\documentclass[%
reprint,
 amsmath,amssymb,
 aps, showkeys,
]{revtex4-1}

\usepackage{graphicx}
\usepackage{dcolumn}
\usepackage{bm}
\usepackage[caption=false]{subfig}

\begin{document}


\title{High efficiency phase flattening based Laguerre-Gauss (LG) spectrometer using variable focus lenses}

\author{Mumtaz Sheikh}
 \email{mumtaz.sheikh@lums.edu.pk}
\author{Haad Yaqub Rathore}
\author{Sohaib Abdul Rehman}%
\affiliation{%
 Department of Physics, School of Science and Engineering, Lahore University of Management Sciences, DHA, Lahore Cantt., Pakistan 54792
}%

\date{\today}

\begin{abstract}
We propose a novel high-efficiency no-moving-parts Laguerre-Gauss (LG) spectrometer using two variable focus lenses and a variable sized pinhole that overcomes the limitations of the classical, projective, phase flattening technique for measuring the Laguerre-Gauss (LG) spectrum of light beams. Simulation results show that the coupling losses are virtually zero and the only losses are ring losses which are mode-dependent but beam waist-independent. Hence, the detection efficiency for all modes is simultaneously the maximum possible irrespective of the beam waist of the LG modes chosen for the decomposition.  The losses can also be easily pre-calibrated to remove the efficiency bias amongst different modes. 
\end{abstract}

\keywords{Paraxial wave optics, singular optics, spatial light modulators, adaptive optics}
\maketitle



Recently, there has been a lot of interest in the use of Laguerre-Gaussian (LG) beams carrying orbital angular momentum (OAM) \cite{Allen:92} for both free-space \cite{Wang:12}, and fiber \cite{Bozinovic:13} communication. Various methods have been proposed to measure the OAM of a single photon \cite{Leach:02,Wei:03,Giovanni:12} as well as the OAM spectrum of light beams \cite{Vasnetsov:03,Berkhout:08,Lavery:11}. The measurement of LG spectrum of an unknown beam, though, remains tricky and direct methods such as projective phase-flattening \cite{Mair:01} and then coupling into a single mode fiber (SMF) have their limitations \cite{Qassim:14}. In this paper, we address some of the limitations of the projective phase-flattening approach using electronically controlled variable focus lenses. Such variable focus lenses have been used in a number of applications \cite{Riza:13}.

The projective phase-flattening approach works by projecting an unknown, incoming beam onto a conjugate Laguerre-Gaussian (LG) mode using a phase spatial light modulator (SLM). A Fourier lens is then used to take the Fourier transform of the resultant field at the phase SLM in the focal plane of the lens. This Fourier-transformed field is then coupled into a single-mode-fiber (SMF). If the input beam contains that particular mode, then the helical phase of the input beam is completely canceled and the Fourier-transformed field has a central bright spot similar to a Gaussian with a ringed intensity pattern around it. The central bright spot can then couple into an SMF as the SMF supports only the $TEM_{{00}}$ mode which is also similar to a Gaussian. The process is repeated for different modes to determine the complete LG spectrum.

The limitations of this approach are highlighted in Ref. \cite{Qassim:14}. Firstly, the detection efficiency into the SMF varies from mode to mode and also with the selected value of beam waist $w_0$ of the LG mode (see Ref \cite{Qassim:14}, Fig. 2). The maximum possible detection efficiency for all the different possible modes is not obtained for one particular value of $w_0$, which implies that all modes cannot optimally couple into the SMF simultaneously. Moreover, for a particular value of $w_0$, the detection efficiency decreases with mode order thus limiting the bandwidth of the measured OAM spectrum. Ideally, the choice of $w_0$ should be such that it gives high enough detection efficiencies for all modes. Secondly, the radial decomposition of the beam depends upon the value of the beam waist $w_0$ chosen for the modes. Optimal choice of the beam waist is the one that gives the minimum number of radial modes. However, that cannot be known \textit{a priori} for an unknown incoming beam. Finally, since the detection efficiency is both mode as well as beam waist-dependent, to determine the complete OAM spectrum, there is a need to post-process the measured data to remove the bias in detection efficiencies.

\begin{figure}[htbp]
\centering
\includegraphics[width=\linewidth]{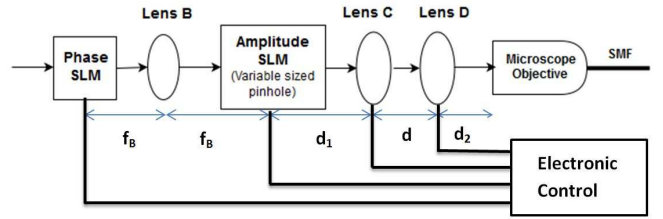}
\caption{Proposed design to measure the LG spectrum of an unknown, incoming beam.}
\label{fig:fig1}
\end{figure}

\begin{figure}[htbp]
\centering
\includegraphics[width=0.79\linewidth]{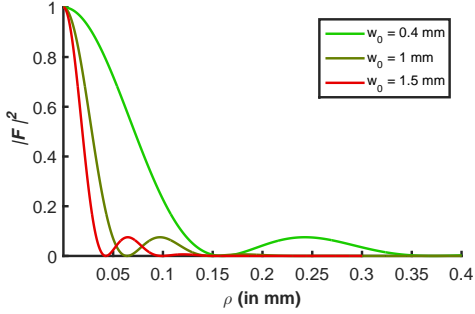}
\caption{Cross-section of the intensity at the focal length of lens B for different beam waists and fixed $p=0$ and $l=4$. The number of radial singularities is not affected by the beam waist, but the size of the central region increases with a decrease in the beam waist of the LG mode. Normalized units are used for all quantities.}
\label{fig:fig2}
\end{figure}

Figure \ref{fig:fig1}
shows our proposed design for measuring the LG spectrum of an unknown, incoming beam. The main idea is to modify the classical projective, phase-flattening approach so that the detection efficiencies are beam waist-independent and are always at their maximum possible values. This is achieved by eliminating coupling losses via the use of two variable focus lenses and a variable-sized pinhole \cite{Sheikh:15}. In Fig. \ref{fig:fig1}, the phase SLM projects the incoming unknown beam on a conjugate LG mode while lens B takes the Fourier transform of the resultant beam exactly as before \cite{Mair:01}. If there is a mode match with the incoming beam, then in the Fourier plane of lens B, the resultant beam would have a central bright spot with a ringed intensity pattern. The fraction of power present in the central region is independent of the beam waist of the LG mode as changing the beam waist only magnifies/de-magnifies the pattern (see Fig. \ref{fig:fig2}).
An amplitude SLM functioning as a variable-sized pinhole selects only the central bright Gaussian-like part of the beam in case of a mode match with the incoming beam. The use of the pinhole is important to ensure that the beam becomes Gaussian-like so that it will stay as a Gaussian on passing through the two variable focus lenses and the microscope objective lens before coupling into the SMF. This Gaussian-like beam is then mode-matched to the SMF using two-variable focus lenses and the microscope objective lens. We show that this system makes the coupling losses into the SMF go to virtually zero for all modes. The reason that the coupling losses go to zero is that not only the transverse field amplitude but also its phase curvature is matched to the fiber mode. Note that no extra intensity is blocked by the pinhole compared to before as the intensity in the outer rings did not couple into the SMF in any case \cite{Qassim:14}. The detection efficiency for all modes then is beam-waist independent and simultaneously the maximum possible.

An LG mode can be characterized by two indices $l$ and $p$, which represent the azimuthal number and radial index of the beam respectively. Different LG modes are mutually orthogonal and form a complete set of solutions to the paraxial wave equation. The mode can be mathematically represented by Eq. \ref{eq:eq1} at the pupil:
\begin{equation}
LG_{p, l}(r,\phi) = \sqrt{\frac{2^{|l|+1}p!}{\pi w_0^2 (p+|l|)!}}
\left(\frac{r}{w_0}\right)^{|l|} e^{-\frac{r^2}{w_0^2}} L_p^{|l|} \left(\frac{2r^2}{w_0^2}\right)
e^{-il\phi}
\label{eq:eq1}
\end{equation}
\noindent
where $r$ is the radial coordinate, $\phi$ is the azimuthal angle,  $w_0$ is the beam waist radius at the pupil and  $L_p^{|l|}(.)$ is the generalized Laguerre polynomial.

This mode is projected on to a possibly different conjugate LG mode $LG_{p',l'}^{*}$ using the phase SLM. The chosen beam waist radius of this conjugate mode gives the $w_0$ for the decomposition of the unknown incoming beam. The resulting field is Fourier-transformed in the focal plane of lens B. The phase SLM is placed in the front focal plane of lens B, so that the field in the back focal plane of lens B is given by a 2-D Fourier transform. The field in the back focal plane of lens B is then given by \cite{Qassim:14}:
\begin{equation}
\mathcal{F}(\rho,\varphi)=\mathcal{FT}[LG_{p, l}(r,\phi) LG_{p', l'}^{*}(r,\phi)]
\label{eq:eq2}
\end{equation}
\noindent
where $\mathcal{FT}$ stands for the 2-D Fourier transform and $\rho$ and $\varphi$ are the transverse coordinates in this plane. If we get a mode match with the incoming beam i.e. $p=p'$ and $l=l'$, the Fourier transform gives a central bright Gaussian-like beam in the back focal plane of lens B. The resultant phase-flattened beam is cylindrically symmetric in the plane of the phase SLM (and remains so throughout the system), therefore we can reduce Eq. \ref{eq:eq2} to the 1-D Hankel transform of order zero \cite{Peatross:15}:
\begin{equation}
\mathcal{F}(\rho,\varphi)=2\pi\frac{e^{i2kf_B}}{i\lambda f_B} \\ \int\limits_{0}^{\infty} \ r|LG_{p, l}|^2J_0\left(\frac{2\pi}{\lambda f_B}r\rho\right) \,dr
\label{eq:eq3}
\end{equation}
\noindent
where $\lambda$ is the wavelength of the light, $k=2\pi /\lambda$, $f_B$ is the focal length of lens B and $J_0(.)$ is the Bessel function of the first kind of order zero. If there is a mode mismatch, we do not get a central maximum at the amplitude SLM and the beam does not couple efficiently into the SMF as our simulations later show. Some examples of what the beam looks like in this plane when there is a mode match are given in Ref. \cite{Qassim:14}, Fig. 1 for different values of $p$ and $l$. Some power moves to the outer rings as both $p$ and $|l|$ increase. The power present in the outer rings cannot couple into the SMF.

The pinhole placed in the back focal plane of lens B, truncates the field at the first zero (see Fig. \ref{fig:fig2}) and allows only the central Gaussian-like region to pass through. The size of the pinhole needs to be variable to cater for different modes and $w_0$'s chosen for the decomposition. Hence, we use an amplitude SLM functioning as this variable sized pinhole. Besides constant parameters such as the wavelength of light $\lambda$ and the focal length of lens B, $f_B$, the radius of this variable-sized pinhole, $s$, completely depends upon the $p$ and $l$ values of the mode under consideration and the $w_0$ chosen for the decomposition and is therefore, set in conjunction with the phase SLM. The field after the amplitude SLM is given by:
\begin{equation}
E_0(\rho,\varphi)=\mathcal{F}(\rho,\varphi)circ(s)
\label{eq:eq4}
\end{equation}
\noindent
where $s$ is a function of $p$, $l$ and $w_0$ and $circ(s)$ is the 2-D circle function which has a value of unity for $\rho < s$ and zero otherwise.

The resulting field $E_0$ then propagates through two variable focus lenses with focal lengths $f_1$ and $f_2$ and to couple into the SMF via the microscope objective lens. Ideal coupling into the SMF depends on matching both the size of the beam and its phase curvature to that of the SMF mode. We, therefore, need to have two independent parameters that we can vary in order to control both the size and phase curvature of the beam. In this case, these two parameters are $f_1$ and $f_2$. The field at the SMF is given by repeated application of the Fresnel diffraction integral.  For example, if the field just after lens C is given by $E_1(\rho',\varphi')$, where $\rho'$ and $\varphi'$ are transverse plane coordinates, then the field $E_2(\rho,\varphi)$ just after lens D, where $\rho$ and $\varphi$ are transverse plane coordinates, is given by \cite{Goodman:05} (in cylindrical coordinates):
\begin{multline}
E_2(\rho,\varphi)= -\frac{ik}{d}e^{ikd}e^{i\frac{k\rho^2}{2d}} e^{-i\frac{k\rho^2}{2f_2}} \\  \left(\int\limits_{0}^{\infty}\rho'E_1(\rho',0)e^{i\frac{k\rho'^2}{2d}}J_0\left(\frac{k\rho\rho'}{d}\right) \,d\rho' \right)
\label{eq:eq5}
\end{multline}
\noindent
where $d$ is the separation between the lenses, $\lambda$ is the wavelength of the light and $k=2\pi /\lambda$. In the same way, the field $E_1$ starting from the field $E_0$ at the pinhole and the field $E_3$ at the SMF starting from the field $E_2$ after lens D may be calculated.

The power coupled into the SMF is given by the overlap integral of the incoming field $E_3(\rho,\varphi)$ with that of the fiber mode which can be approximated by a Gaussian. Expressing this integral in cylindrical coordinates and making use of the inherent cylindrical symmetry due to phase-flattening, the coupling efficiency is given by \cite{Molina-Terriza:07}:
\begin{equation}
\eta_c = \frac{2}{\pi \sigma^2}\frac{\left|\int\limits_{0}^{\infty}\rho E_3^*(\rho,0)e^{-\frac{\rho^2}{\sigma^2}}\,d\rho\right|^2}{{\int\limits_{0}^{\infty}\rho\left|E_3(\rho,0)\right|^2}\,d\rho}
\label{eq:eq6}
\end{equation}
\noindent
where $\sigma$ is the beam waist radius of the SMF mode.

To find the values of the focal lengths, $f_1$ and $f_2$, of the two variable focus lenses in order to maximally couple the field $E_1$ after the pinhole into the SMF, we use the ABCD matrix approach \cite{Kogelnik:66}. The field $E_1$ can be approximated with a Gaussian with radius $w_1$ taken to be the point where the field drops to $1/e$ of its maximum value. Just like the size of the pinhole $s$, $w_1$ is also completely determined by $p$, $l$ and $w_0$. The $q$-parameter \cite{Kogelnik:66} of this input Gaussian is then given by:
\begin{equation}
q_1=\frac{ikw_1^2}{2}
\label{eq:eq7}
\end{equation}
The ABCD matrix for the system is given by:
\begin{multline}
\begin{bmatrix}
A & B \\
C & D
\end{bmatrix}
= \begin{bmatrix}
1 & f_m \\
0 & 1
\end{bmatrix}
\begin{bmatrix}
1 & 0 \\
\frac{-1}{f_m} & 1
\end{bmatrix}
\begin{bmatrix}
1 & d_2 \\
0 & 1
\end{bmatrix}
\begin{bmatrix}
1 & 0 \\
\frac{-1}{f_2} & 1
\end{bmatrix}  \\
\begin{bmatrix}
1 & d \\
0 & 1
\end{bmatrix}
\begin{bmatrix}
1 & 0 \\
\frac{-1}{f_1} & 1
\end{bmatrix}
\begin{bmatrix}
1 & d_1 \\
0 & 1
\end{bmatrix}
\label{eq:eq8}
\end{multline}
\noindent
where $d_1$ is the distance from the pinhole to lens C, $d_2$ is the distance from lens D to the microscope objective and $f_m$ is the focal length of the microscope objective used to couple the beam into the SMF. The $q$-parameter of the beam at the fiber needs to conform to the beam waist radius of the fiber mode, $\sigma$, and is then given by:
\begin{equation}
q_2=\frac{ik\sigma^2}{2}=\frac{Aq_1+B}{Cq_1+D}
\label{eq:eq9}
\end{equation}
The only two unknowns in Eq. \ref{eq:eq9} are $f_1$ and $f_2$. By equating the real and imaginary parts on both sides of Eq. \ref{eq:eq9}, we can find the values of $f_1$ and $f_2$ required for ideal coupling. Some design considerations and an experimental demonstration for such a two-lens coupling system are given in \cite{Qasim:15}.

Simulations are carried out to test the proposed Fig. \ref{fig:fig1} design for different values of $w_0$, $l$ and $p$ and calculating the detection efficiencies $\eta$ for the optimal choice of parameters $s$, $f_1$ and $f_2$. The values of the fixed parameters of the system are taken as follows: $\lambda=500$ nm, $f_B=50$ cm, $d_1= 10$ cm, $d=20$ cm, $d_2=10$ cm, $f_m = 10$ mm and $\sigma = 5$ $\mu$m. 

\begin{figure}[htbp]
\centering
\subfloat[]{\label{fig:fig3a}%
  \includegraphics[width=0.5\linewidth]{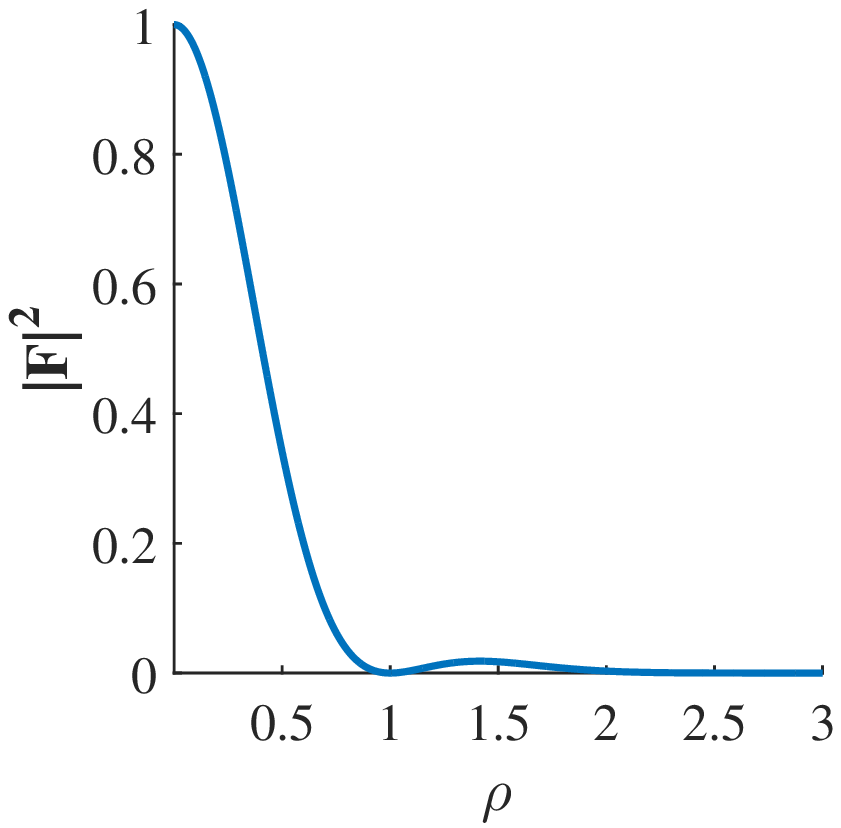}%
}
\subfloat[]{\label{fig:fig3b}%
  \includegraphics[width=0.5\linewidth]{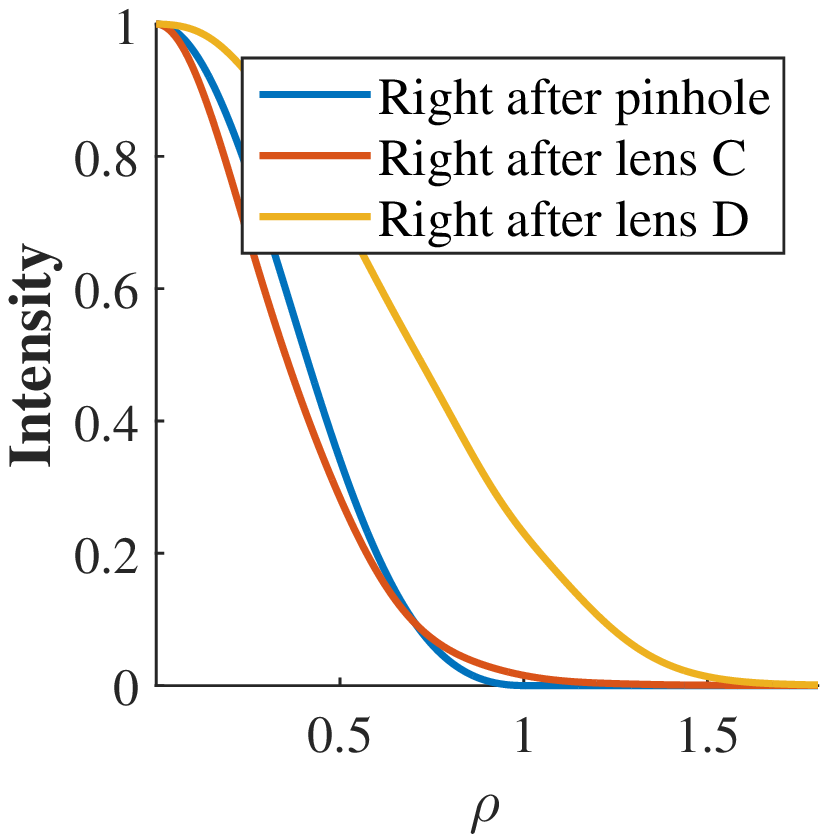}%
}\\
\subfloat[]{\label{fig:fig3c}%
  \includegraphics[width=0.5\linewidth]{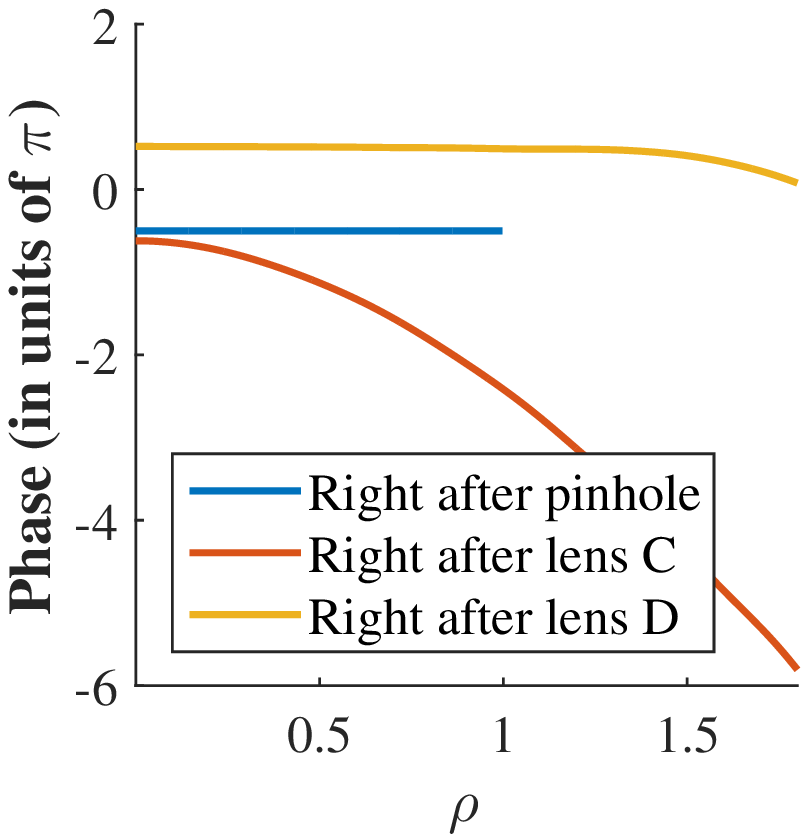}%
}
\subfloat[]{\label{fig:fig3d}
  \includegraphics[width=0.5\linewidth]{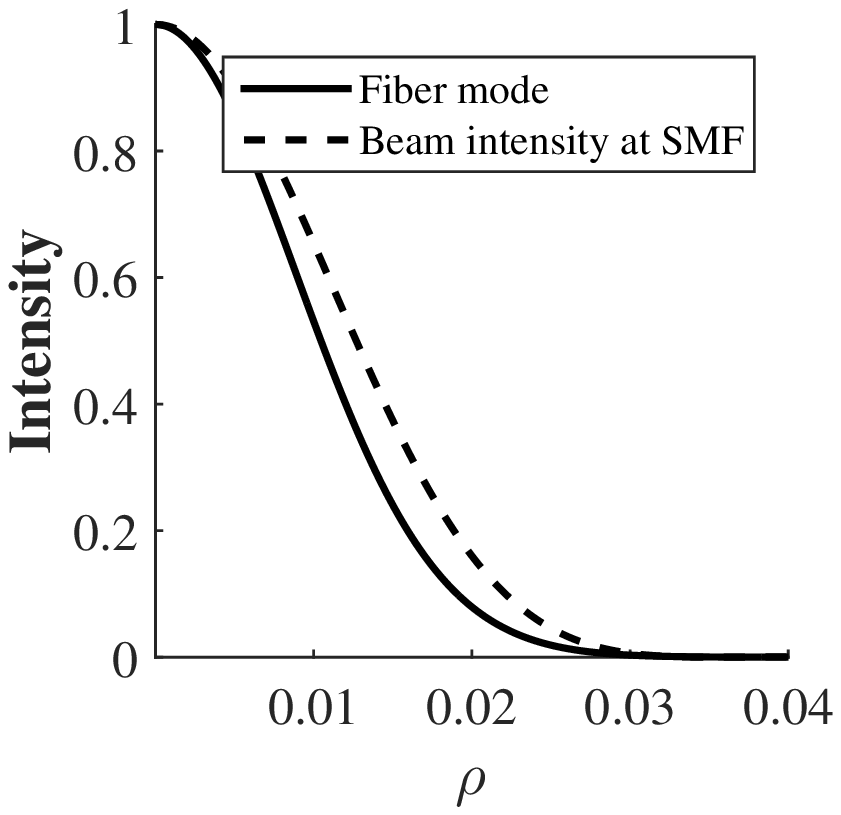}
}\\
\subfloat[]{\label{fig:fig3e}
  \includegraphics[width=0.5\linewidth]{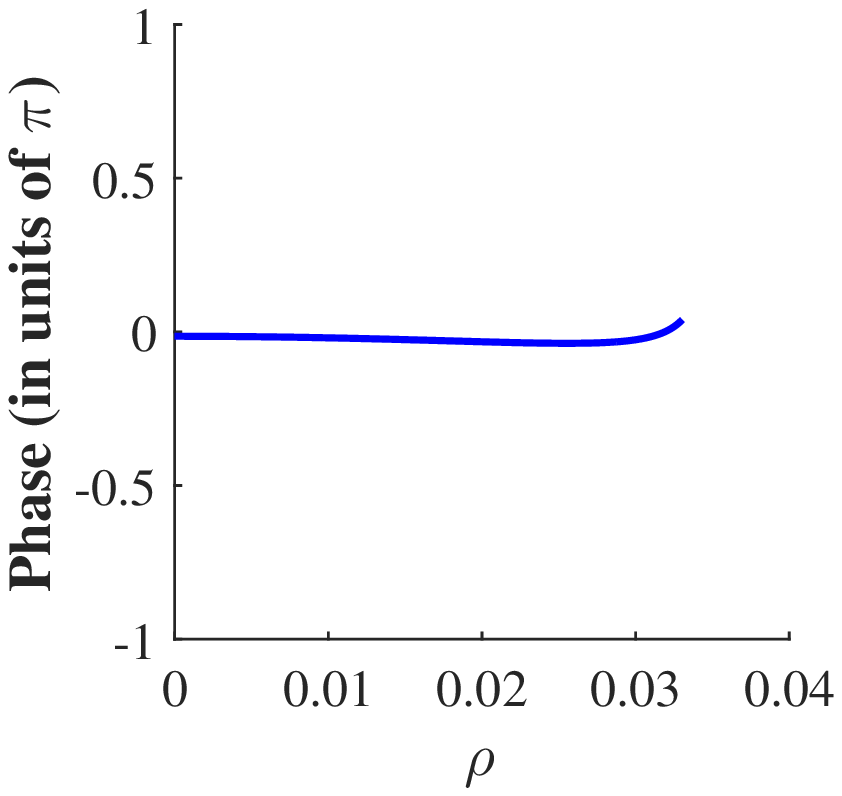}
}
\caption{Propagation of $p=0$, $l=1$ mode through the system. (a) Mode intensity just before the amplitude SLM. (b) Mode intensity at different planes through the system. (c) Phase profile (in units of $\pi$) at different planes through the system. (d) Overlap of the mode intensity at the SMF with the fiber mode. (e) Phase profile of the mode at the SMF.}
\label{fig:fig3}
\end{figure}

Figure \ref{fig:fig3} 
shows the mode cross-section results for the $p=0$, $l=1$ beam when $w_0=0.40$ mm at different planes along the z-axis. We have normalized the $\rho$ axis by $a_0=\frac{\sqrt{2}\lambda f_B}{\pi w_0}$, the natural scaling factor in the far field of lens B \cite{Qassim:14}. In case of a mode match at the phase SLM with the incoming beam, i.e. $p=p'$ and $l=l'$, the mode just before the amplitude SLM is given in Fig. \ref{fig:fig3a} as determined from Eq. \ref{eq:eq3}. The first zero of this mode is at $\rho=a_0=0.282$ mm. The radius of the variable-sized pinhole on the amplitude SLM is therefore set to $s=0.282$ mm in Eq. \ref{eq:eq4}. Next, the values of $f_1$ and $f_2$ are determined using the ABCD matrix approach given in Eqs. \ref{eq:eq7}-\ref{eq:eq9}. The value of $w_1$, where the intensity drops to $1/e^2$ of its maximum value, in Eq. \ref{eq:eq7} is found to be $w_1=0.185$ mm. Using Eq. \ref{eq:eq7} and \ref{eq:eq8}, the values of $f_1$ and $f_2$ are determined to be $f_1=7.31$ cm and $f_2=12.5$ cm. Fig. \ref{fig:fig3b} shows the intensity profile of this truncated beam as it progresses through the system (till the microscope objective), obtained using three applications of the Fresnel diffraction integral, such as the one given in Eq. \ref{eq:eq5}, using these values of $f_1$ and $f_2$. The corresponding phases are shown in Fig. \ref{fig:fig3c}. Note that since the variable-sized pinhole truncates the beam at $s=a_0$, the phase of the beam right after the pinhole is undefined for all $\rho > a_0$. Fig. \ref{fig:fig3d} shows the intensity profile of the mode, after passing through the entire system (including the microscope objective), at the plane of the SMF overlapped with the target mode of the SMF. Similarly, Fig. \ref{fig:fig3e} shows the phase profile of the mode at the plane of the SMF. This is again achieved by another application of the Fresnel diffraction integral. Notice that the phase of the transverse field at the plane of the SMF is virtually constant. Consequently, the coupling efficiency between the two, as given by Eq. \ref{eq:eq6}, is found to be $\eta_c=0.98$. The only other loss in the system is due to truncation of the beam at the pinhole. The fraction of power transmitted at the pinhole, $\eta_p$, is found to be $\eta_p=0.86$. The detection efficiency for the $p=0$, $l=1$ mode is hence, found to be $\eta=\eta_p\eta_c=0.85$. Note that this is exactly equal to the corresponding maximum theoretical value of the detection efficiency for $p=0$, $l=1$ mode given in Ref. \cite{Qassim:14}. If we choose a different value of $w_0$ for the decomposition of the incoming beam, such as $w_0 = 1$ mm, then the corresponding optimal parameters are found to be $s=a_0=0.113$ mm, $f_1=4.93$ cm and $f_2 = 11.9$ cm whereas the detection efficiency is again found to $\eta=\eta_p\eta_c=0.86\times 0.99=0.85$. Note that the fraction of power contained in the central bright spot, $\eta_p$, remains the same for both cases while the coupling efficiency, $\eta_c$, varies only slightly. Therefore, as predicted, the overall efficiency remains the maximum possible irrespective of the choice of $w_0$.

\begin{figure}[htbp]
\centering
\subfloat[]{\label{fig:fig4a}%
  \includegraphics[width=0.5\linewidth]{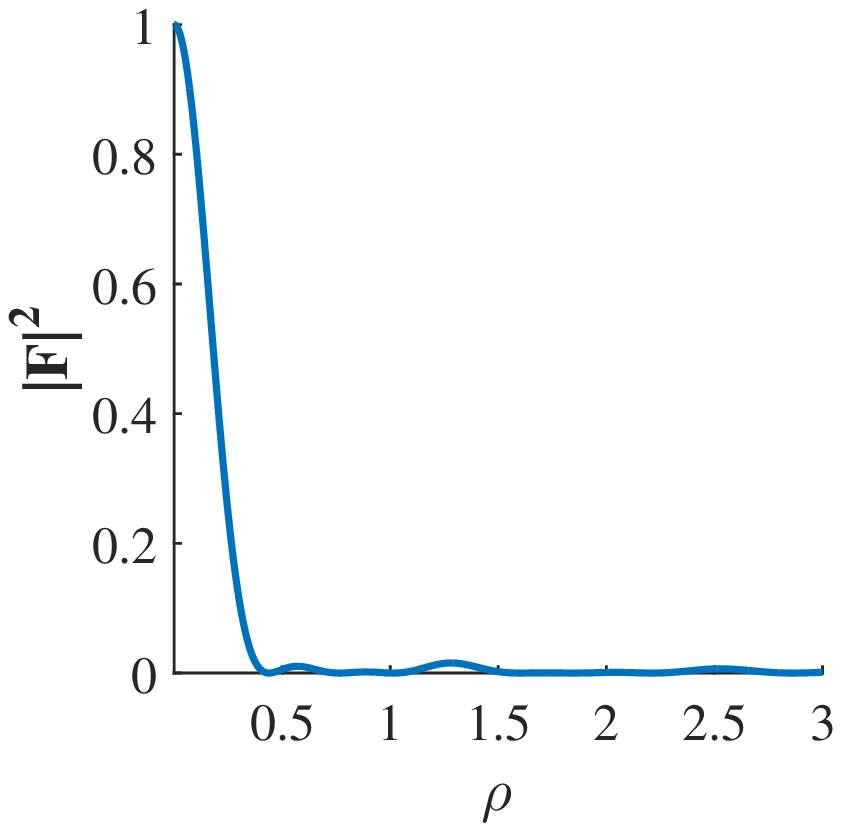}%
}
\subfloat[]{\label{fig:fig4b}%
  \includegraphics[width=0.5\linewidth]{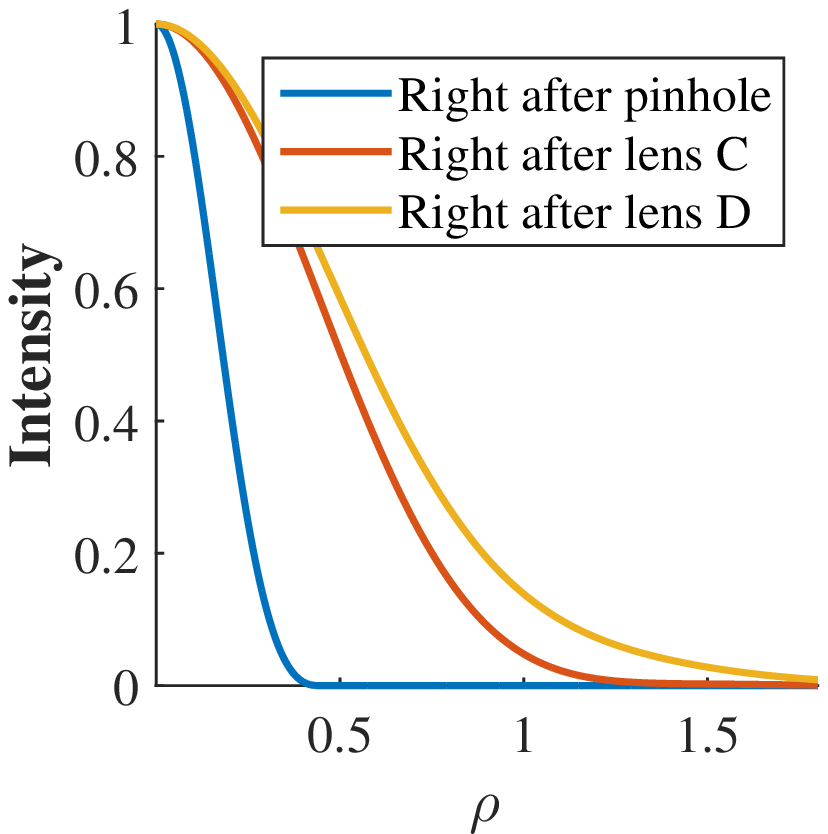}%
}\\
\subfloat[]{\label{fig:fig4c}%
  \includegraphics[width=0.5\linewidth]{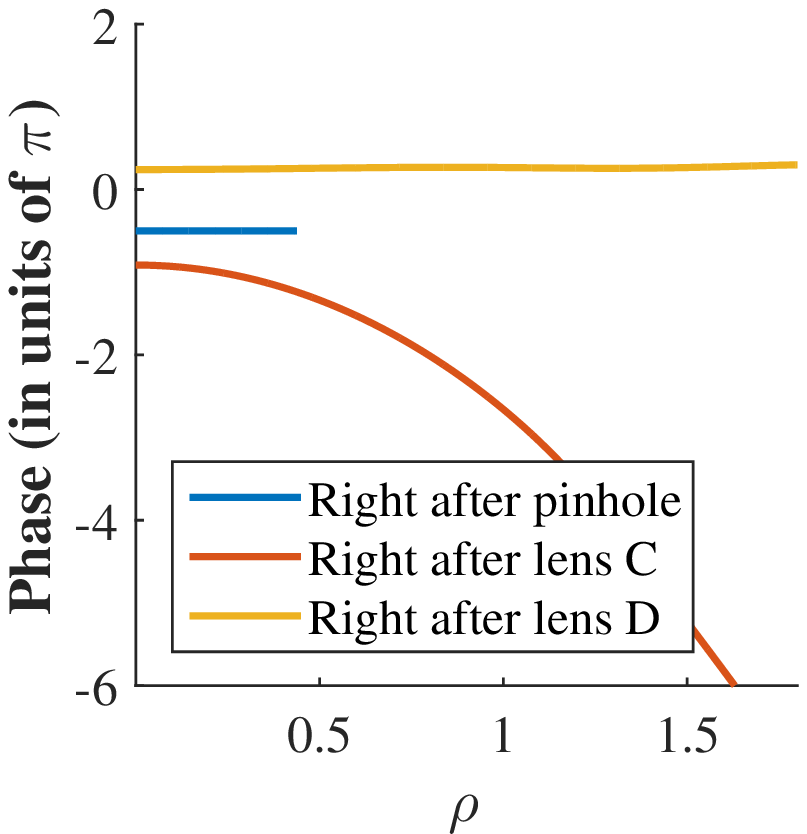}%
}
\subfloat[]{\label{fig:fig4d}
  \includegraphics[width=0.5\linewidth]{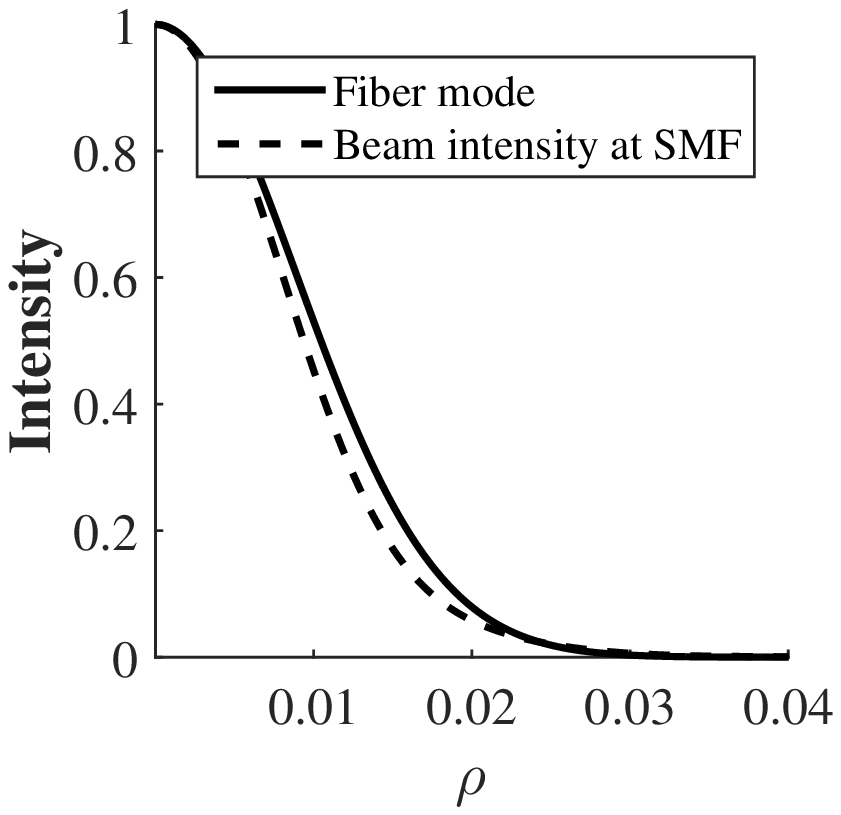}%
}\\
\subfloat[]{\label{fig:fig4e}
  \includegraphics[width=0.5\linewidth]{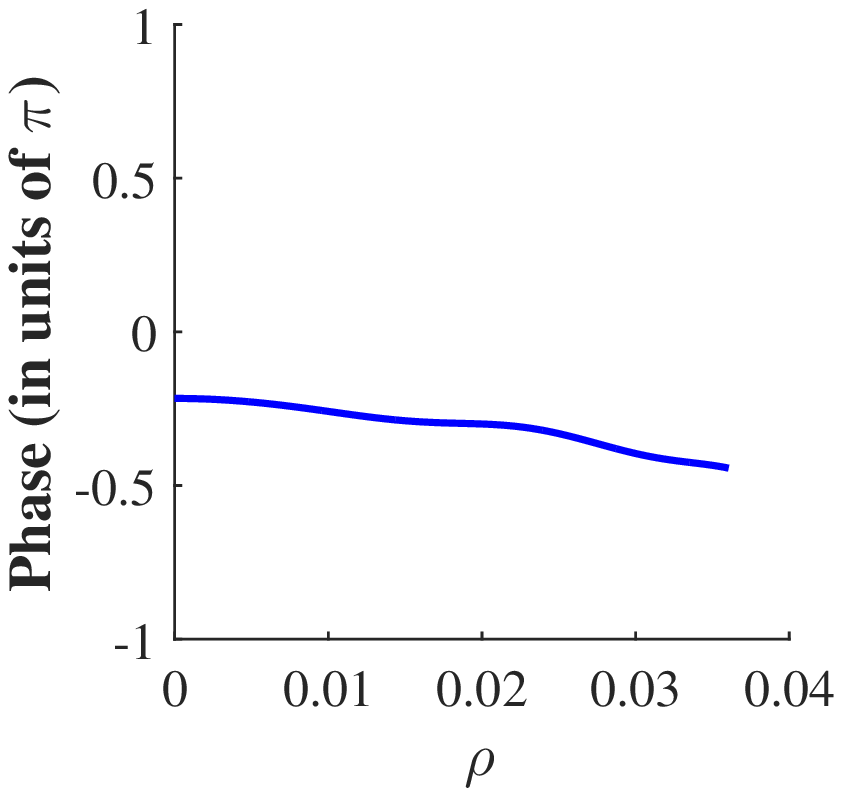}
}
\caption{Propagation of $p=2$, $l=5$ mode through the system. (a) Mode intensity just before the amplitude SLM. (b) Mode intensity at different planes through the system. (c) Phase profile (in units of $\pi$) at different planes through the system. (d) Overlap of the mode intensity at the SMF with the fiber mode. (e) Phase profile of the mode at the SMF.}
\label{fig:fig4}
\end{figure}

The simulation results for a higher-order mode $p=2$, $l=5$ and $w_0=0.40$ mm mode are shown in Figure \ref{fig:fig4}. As before, the $\rho$ axis has been normalized by $a_0$.
In this case, the optimal choice of parameters are found to be $s=0.44a_0=0.124$ mm, $f_1=4.94$ cm and $f_2 = 12.3$ cm whereas the detection efficiency is found to $\eta=\eta_p\eta_c=0.60\times 0.98=0.59$. The coupling losses are again virtually zero as the transverse field phase is virtually constant at the plane of the SMF. The detection efficiency, however, is much lower compared to the $p=0$, $l=1$ mode due to more power moving to outer rings.

Table \ref{tab:tab1} 
\begin{table}[htbp]
\centering
\caption{\bf Optimal parameters and detection efficiencies for different modes}
\begin{tabular}{cccccccc}
\hline
Mode & $w_0$ & $s$ & $f_1$  & $f_2$ & $\eta_p$ & $\eta_c$ & $\eta$ \\
(p,l) & (mm) & (mm) & (cm) & (cm) & & & \\
\hline
$(0,0)$ & $0.6$ & $0.375$ & 7.37 & 12.5 & 1.00 & 1.00 & 1.00 \\
$(0,0)$ & $0.8$ & $0.281$ & 6.10 & 13.2 & 1.00 & 1.00 & 1.00 \\
$(0,4)$ & $0.6$ & $0.106$ & 4.93 & 12.0 & 0.59 & 0.97 & 0.58 \\
$(0,4)$ & $0.8$ & $0.080$ & 5.04 & 10.6 & 0.59 & 0.98 & 0.58 \\
$(1,4)$ & $0.6$ & $0.096$ & 4.96 & 11.3 & 0.63 & 0.98 & 0.62 \\
$(1,4)$ & $0.8$ & $0.072$ & 5.15 & 9.88 & 0.63 & 0.98 & 0.62 \\
$(2,6)$ & $0.6$ & $0.077$ & 5.10 & 10.2 & 0.58 & 0.98 & 0.57 \\
$(3,2)$ & $0.6$ & $0.096$ & 5.01 & 10.8 & 0.61 & 0.99 & 0.60 \\
\hline
\end{tabular}
\label{tab:tab1}
\end{table}
shows the optimal values of the various parameters along with the corresponding detection efficiencies for some selected modes. It shows that the fraction of power in the central bright region, $\eta_p$, indeed remains constant with $w_0$ for different modes $(p,l)$. It also shows that the coupling efficiency, $\eta_c$, remains above $0.97$ in all cases.

\begin{figure}[htbp]
\centering
\subfloat[]{\label{fig:fig5a}%
  \includegraphics[width=0.5\linewidth]{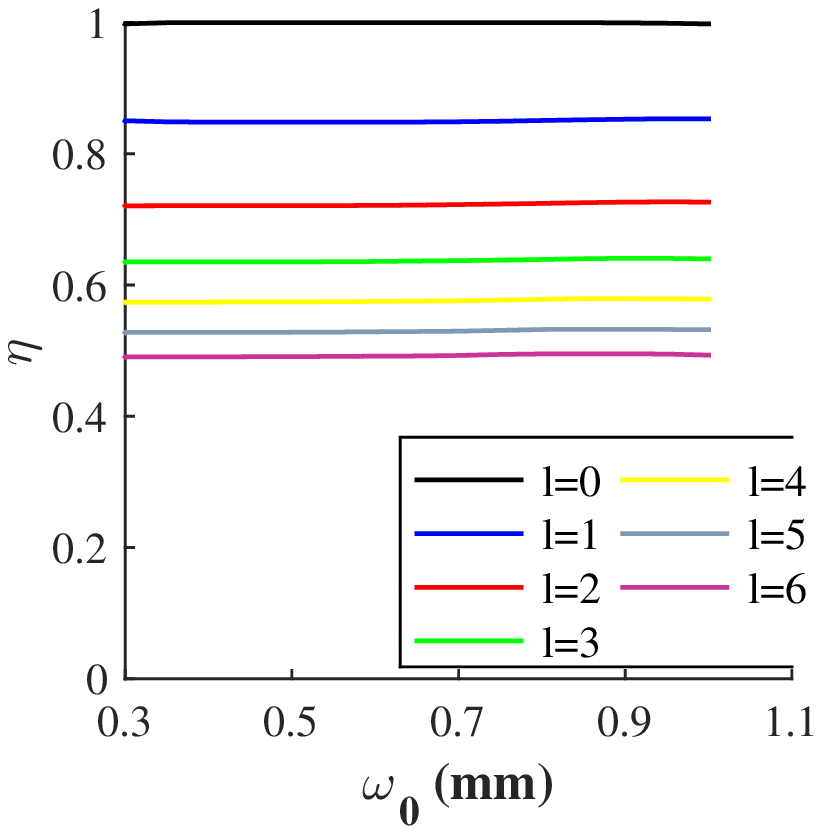}%
}
\subfloat[]{\label{fig:fig5b}%
  \includegraphics[width=0.5\linewidth]{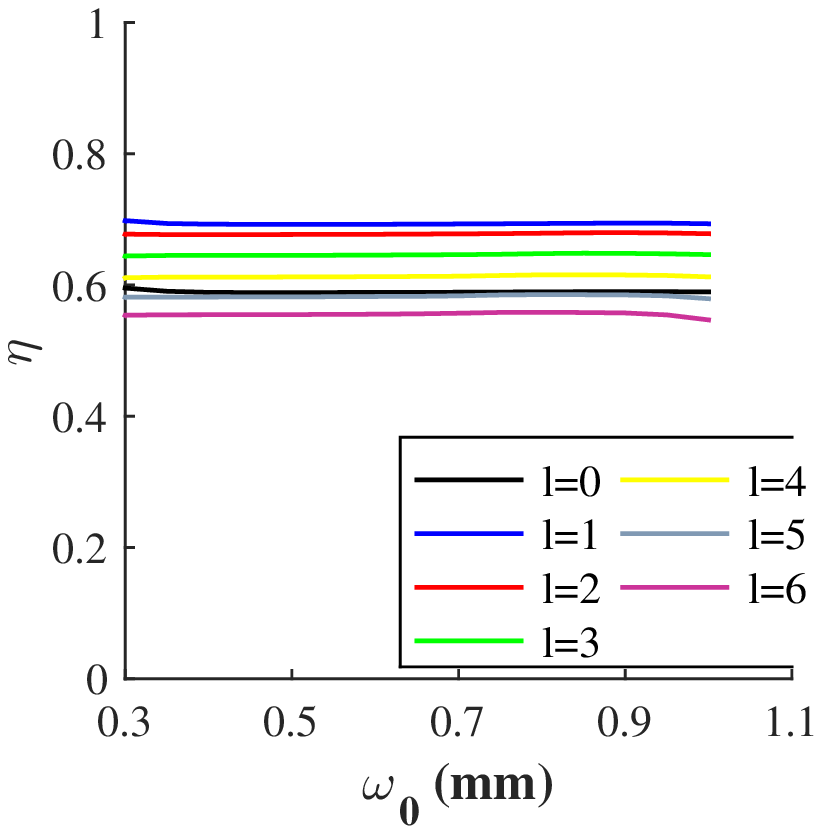}%
}\\
\subfloat[]{\label{fig:fig5c}%
  \includegraphics[width=0.5\linewidth]{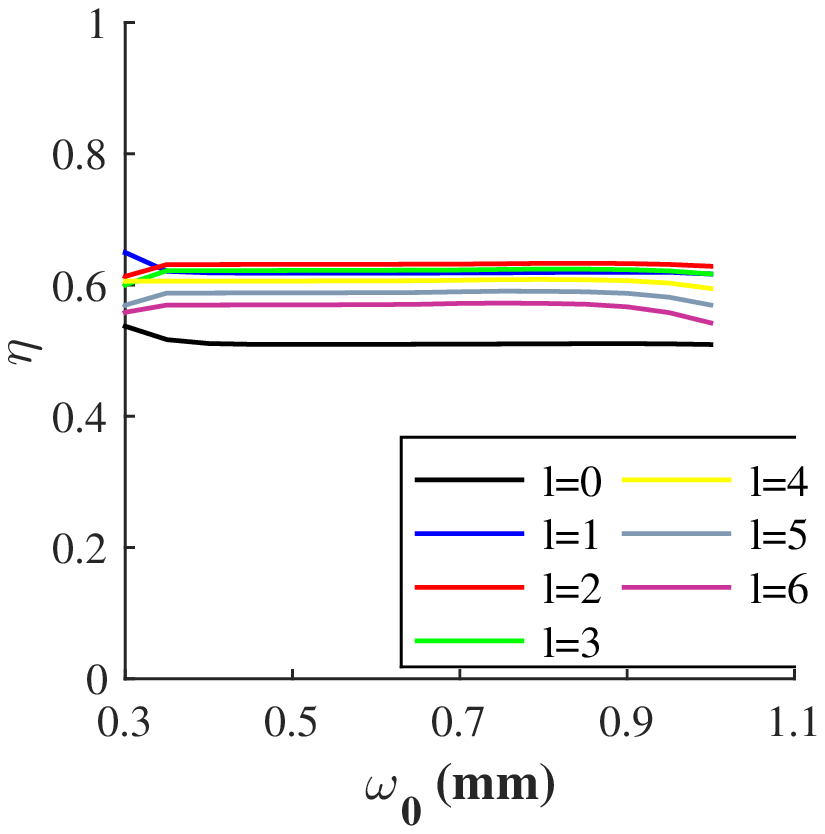}%
}
\caption{Detection efficiency, $\eta$, for different modes against $w_0$, the beam waist chosen for the decomposition (a) $p=0$ (b) $p=1$ (c) $p=2$.}
\label{fig:fig5}
\end{figure}

Figure \ref{fig:fig5} 
shows that the detection efficiency, $\eta$, for each mode remains virtually constant, as predicted, for different values of the beam waist $w_0$, chosen for the decomposition. The actual value of $\eta$ remains the maximum possible for all modes simultaneously including higher-order modes. Thus, the efficiency curves of Ref. \cite{Qassim:14}, Fig. 2 are effectively flattened at their peak values. The efficiency bias between different modes, therefore, remains constant which may be easily pre-calibrated.

\begin{figure}[htbp]
\centering
\subfloat[]{\label{fig:fig6a}%
  \includegraphics[width=0.5\linewidth]{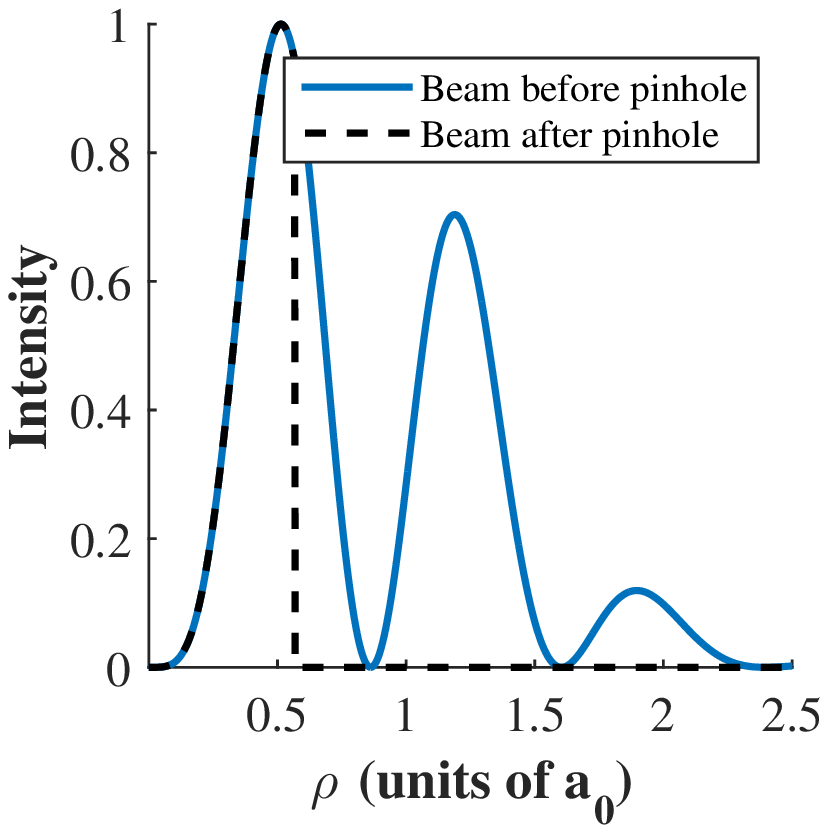}%
}
\subfloat[]{\label{fig:fig6b}%
  \includegraphics[width=0.5\linewidth]{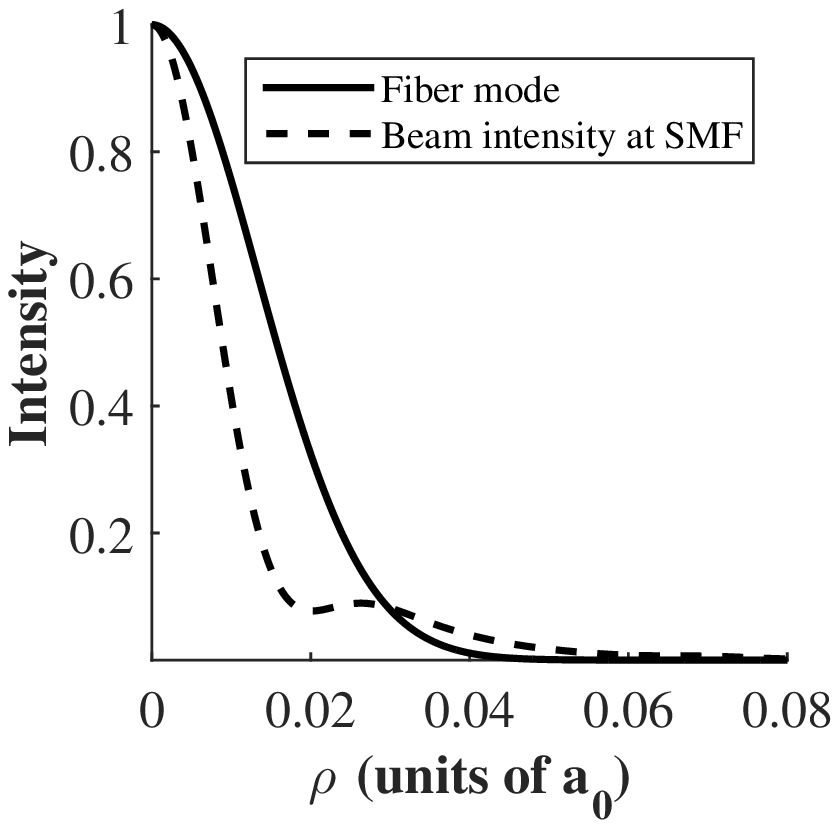}%
}\\
\subfloat[]{\label{fig:fig6c}
   \includegraphics[width=0.5\linewidth]{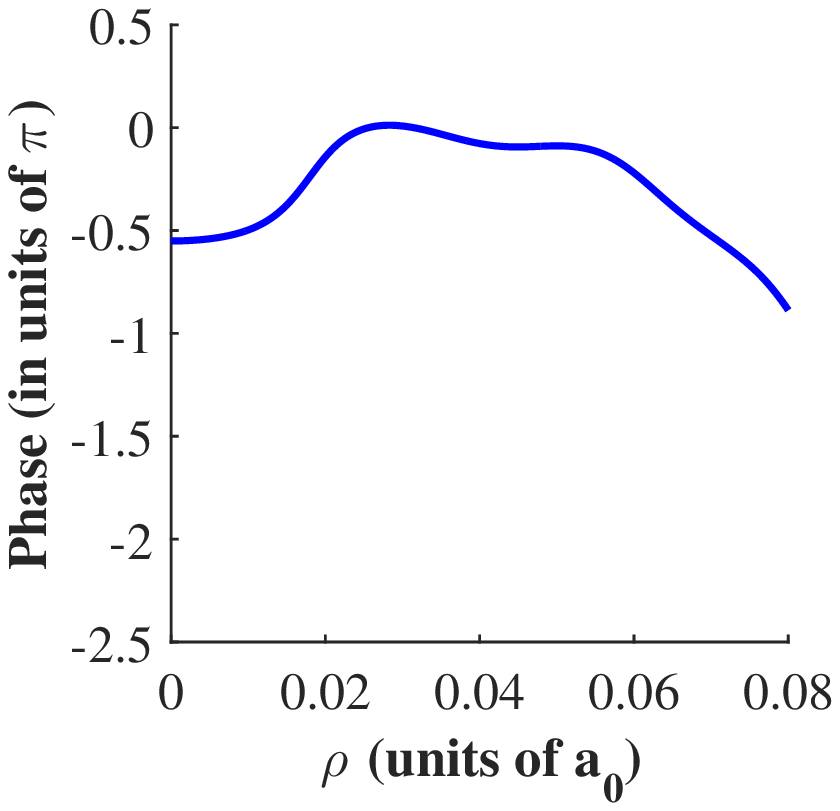}
}
\caption{The case when an incoming mode $(p,l)=(1,4)$ mode is projected onto a $(p',l')=(0,4)$ mode. (a) Intensity profile before and after truncation at the variable-sized pinhole. (b) Overlap of the mode intensity at the SMF with the fiber mode. (c) Phase profile of the mode at the SMF.}
\label{fig:fig6}
\end{figure}

If there is a mode mismatch between the incoming beam and the conjugate LG mode projected on the phase SLM, the detection efficiency is virtually zero. For example, an incoming mode $(p,l) = (1,4)$ with beam waist $0.6$ mm is projected onto the conjugate mode $(p',l') = (0,4)$ with $w_0$ set at $0.6$ mm and the corresponding parameters selected from Table \ref{tab:tab1} as $s=0.106$ mm, $f_1=4.93$ cm and $f_2=12.0$ cm. The beam at the SMF plane after passing through the system overlapped with the target SMF mode is shown in Figure \ref{fig:fig6}. 
Figure \ref{fig:fig6a} compares the beam intensity just before the pinhole with the intensity just after the pinhole. In this case, the fraction of the power transmitted through the pinhole is $\eta_p = 0.18$, which causes the detection efficiency to fall down to $\eta = \eta_p\eta_c = 0.18\times 0.45 = 0.08$. It can therefore be concluded that the cross-talk between different modes is virtually non-existent. It is interesting to note that in Figure \ref{fig:fig6c}, the phase of the beam at the SMF varies by more than $\pi$, hence causing the coupling efficiency, $\eta_c$, to fall as well. This is in contrast to the mode-match situations of Figures \ref{fig:fig3e} and \ref{fig:fig4e} where the phase curvature was almost flat and $\eta_c$ was as high as $0.98$. 

We have shown that in a projective LG measurement, maximum possible detection efficiency can be achieved for all the modes simultaneously using our proposed LG spectrometer. In particular, higher-order OAM modes can also be detected with high efficiency comparable to lower order modes. Future work would focus on an experimental demonstration of the proposed design. 

\bibliography{LGspectbib}

\begin{thebibliography}{18}%
\makeatletter
\providecommand \@ifxundefined [1]{%
 \@ifx{#1\undefined}
}%
\providecommand \@ifnum [1]{%
 \ifnum #1\expandafter \@firstoftwo
 \else \expandafter \@secondoftwo
 \fi
}%
\providecommand \@ifx [1]{%
 \ifx #1\expandafter \@firstoftwo
 \else \expandafter \@secondoftwo
 \fi
}%
\providecommand \natexlab [1]{#1}%
\providecommand \enquote  [1]{``#1''}%
\providecommand \bibnamefont  [1]{#1}%
\providecommand \bibfnamefont [1]{#1}%
\providecommand \citenamefont [1]{#1}%
\providecommand \href@noop [0]{\@secondoftwo}%
\providecommand \href [0]{\begingroup \@sanitize@url \@href}%
\providecommand \@href[1]{\@@startlink{#1}\@@href}%
\providecommand \@@href[1]{\endgroup#1\@@endlink}%
\providecommand \@sanitize@url [0]{\catcode `\\12\catcode `\$12\catcode
  `\&12\catcode `\#12\catcode `\^12\catcode `\_12\catcode `\%12\relax}%
\providecommand \@@startlink[1]{}%
\providecommand \@@endlink[0]{}%
\providecommand \url  [0]{\begingroup\@sanitize@url \@url }%
\providecommand \@url [1]{\endgroup\@href {#1}{\urlprefix }}%
\providecommand \urlprefix  [0]{URL }%
\providecommand \Eprint [0]{\href }%
\providecommand \doibase [0]{http://dx.doi.org/}%
\providecommand \selectlanguage [0]{\@gobble}%
\providecommand \bibinfo  [0]{\@secondoftwo}%
\providecommand \bibfield  [0]{\@secondoftwo}%
\providecommand \translation [1]{[#1]}%
\providecommand \BibitemOpen [0]{}%
\providecommand \bibitemStop [0]{}%
\providecommand \bibitemNoStop [0]{.\EOS\space}%
\providecommand \EOS [0]{\spacefactor3000\relax}%
\providecommand \BibitemShut  [1]{\csname bibitem#1\endcsname}%
\let\auto@bib@innerbib\@empty
\bibitem [{\citenamefont {Allen}\ \emph {et~al.}(1992)\citenamefont {Allen},
  \citenamefont {Beijersbergen}, \citenamefont {Spreeuw},\ and\ \citenamefont
  {Woerdman}}]{Allen:92}%
  \BibitemOpen
  \bibfield  {author} {\bibinfo {author} {\bibfnamefont {L.}~\bibnamefont
  {Allen}}, \bibinfo {author} {\bibfnamefont {M.~W.}\ \bibnamefont
  {Beijersbergen}}, \bibinfo {author} {\bibfnamefont {R.~J.~C.}\ \bibnamefont
  {Spreeuw}}, \ and\ \bibinfo {author} {\bibfnamefont {J.~P.}\ \bibnamefont
  {Woerdman}},\ }\href {\doibase 10.1103/PhysRevA.45.8185} {\bibfield
  {journal} {\bibinfo  {journal} {Phys. Rev. A}\ }\textbf {\bibinfo {volume}
  {45}},\ \bibinfo {pages} {8185} (\bibinfo {year} {1992})}\BibitemShut
  {NoStop}%
\bibitem [{\citenamefont {Wang}\ \emph {et~al.}(2014)\citenamefont {Wang},
  \citenamefont {Yang}, \citenamefont {Fazal}, \citenamefont {Ahmed},
  \citenamefont {Yan}, \citenamefont {Huang}, \citenamefont {Ren},
  \citenamefont {Yue}, \citenamefont {Dolinar}, \citenamefont {Tur},\ and\
  \citenamefont {Willner}}]{Wang:12}%
  \BibitemOpen
  \bibfield  {author} {\bibinfo {author} {\bibfnamefont {J.}~\bibnamefont
  {Wang}}, \bibinfo {author} {\bibfnamefont {J.-Y.}\ \bibnamefont {Yang}},
  \bibinfo {author} {\bibfnamefont {I.~M.}\ \bibnamefont {Fazal}}, \bibinfo
  {author} {\bibfnamefont {N.}~\bibnamefont {Ahmed}}, \bibinfo {author}
  {\bibfnamefont {Y.}~\bibnamefont {Yan}}, \bibinfo {author} {\bibfnamefont
  {H.}~\bibnamefont {Huang}}, \bibinfo {author} {\bibfnamefont
  {Y.}~\bibnamefont {Ren}}, \bibinfo {author} {\bibfnamefont {Y.}~\bibnamefont
  {Yue}}, \bibinfo {author} {\bibfnamefont {S.}~\bibnamefont {Dolinar}},
  \bibinfo {author} {\bibfnamefont {M.}~\bibnamefont {Tur}}, \ and\ \bibinfo
  {author} {\bibfnamefont {A.~E.}\ \bibnamefont {Willner}},\ }\href {\doibase
  10.1038/nphoton.2012.138} {\bibfield  {journal} {\bibinfo  {journal} {Nature
  Photon.}\ }\textbf {\bibinfo {volume} {6}},\ \bibinfo {pages} {488} (\bibinfo
  {year} {2014})}\BibitemShut {NoStop}%
\bibitem [{\citenamefont {Bozinovic}\ \emph {et~al.}(2013)\citenamefont
  {Bozinovic}, \citenamefont {Yue}, \citenamefont {Ren}, \citenamefont {Tur},
  \citenamefont {Kristensen}, \citenamefont {Huang}, \citenamefont {Willner},\
  and\ \citenamefont {Ramachandran}}]{Bozinovic:13}%
  \BibitemOpen
  \bibfield  {author} {\bibinfo {author} {\bibfnamefont {N.}~\bibnamefont
  {Bozinovic}}, \bibinfo {author} {\bibfnamefont {Y.}~\bibnamefont {Yue}},
  \bibinfo {author} {\bibfnamefont {Y.}~\bibnamefont {Ren}}, \bibinfo {author}
  {\bibfnamefont {M.}~\bibnamefont {Tur}}, \bibinfo {author} {\bibfnamefont
  {P.}~\bibnamefont {Kristensen}}, \bibinfo {author} {\bibfnamefont
  {H.}~\bibnamefont {Huang}}, \bibinfo {author} {\bibfnamefont {A.~E.}\
  \bibnamefont {Willner}}, \ and\ \bibinfo {author} {\bibfnamefont
  {S.}~\bibnamefont {Ramachandran}},\ }\href {\doibase 10.1126/science.1237861}
  {\bibfield  {journal} {\bibinfo  {journal} {Science}\ }\textbf {\bibinfo
  {volume} {340}},\ \bibinfo {pages} {1545} (\bibinfo {year} {2013})},\ \Eprint
  {http://arxiv.org/abs/http://www.sciencemag.org/content/340/6140/1545.full.pdf}
  {http://www.sciencemag.org/content/340/6140/1545.full.pdf} \BibitemShut
  {NoStop}%
\bibitem [{\citenamefont {Leach}\ \emph {et~al.}(2002)\citenamefont {Leach},
  \citenamefont {Padgett}, \citenamefont {Barnett}, \citenamefont
  {Franke-Arnold},\ and\ \citenamefont {Courtial}}]{Leach:02}%
  \BibitemOpen
  \bibfield  {author} {\bibinfo {author} {\bibfnamefont {J.}~\bibnamefont
  {Leach}}, \bibinfo {author} {\bibfnamefont {M.~J.}\ \bibnamefont {Padgett}},
  \bibinfo {author} {\bibfnamefont {S.~M.}\ \bibnamefont {Barnett}}, \bibinfo
  {author} {\bibfnamefont {S.}~\bibnamefont {Franke-Arnold}}, \ and\ \bibinfo
  {author} {\bibfnamefont {J.}~\bibnamefont {Courtial}},\ }\href {\doibase
  10.1103/PhysRevLett.88.257901} {\bibfield  {journal} {\bibinfo  {journal}
  {Phys. Rev. Lett.}\ }\textbf {\bibinfo {volume} {88}},\ \bibinfo {pages}
  {257901} (\bibinfo {year} {2002})}\BibitemShut {NoStop}%
\bibitem [{\citenamefont {Wei}\ \emph {et~al.}(2003)\citenamefont {Wei},
  \citenamefont {Xue}, \citenamefont {Leach}, \citenamefont {Padgett},
  \citenamefont {Barnett}, \citenamefont {Franke-Arnold}, \citenamefont {Yao},\
  and\ \citenamefont {Courtial}}]{Wei:03}%
  \BibitemOpen
  \bibfield  {author} {\bibinfo {author} {\bibfnamefont {H.}~\bibnamefont
  {Wei}}, \bibinfo {author} {\bibfnamefont {X.}~\bibnamefont {Xue}}, \bibinfo
  {author} {\bibfnamefont {J.}~\bibnamefont {Leach}}, \bibinfo {author}
  {\bibfnamefont {M.~J.}\ \bibnamefont {Padgett}}, \bibinfo {author}
  {\bibfnamefont {S.~M.}\ \bibnamefont {Barnett}}, \bibinfo {author}
  {\bibfnamefont {S.}~\bibnamefont {Franke-Arnold}}, \bibinfo {author}
  {\bibfnamefont {E.}~\bibnamefont {Yao}}, \ and\ \bibinfo {author}
  {\bibfnamefont {J.}~\bibnamefont {Courtial}},\ }\href {\doibase
  http://dx.doi.org/10.1016/S0030-4018(03)01619-5} {\bibfield  {journal}
  {\bibinfo  {journal} {Optics Communications}\ }\textbf {\bibinfo {volume}
  {223}},\ \bibinfo {pages} {117 } (\bibinfo {year} {2003})}\BibitemShut
  {NoStop}%
\bibitem [{\citenamefont {Giovannini}\ \emph {et~al.}(2012)\citenamefont
  {Giovannini}, \citenamefont {Miatto}, \citenamefont {Romero}, \citenamefont
  {Barnett}, \citenamefont {Woerdman},\ and\ \citenamefont
  {Padgett}}]{Giovanni:12}%
  \BibitemOpen
  \bibfield  {author} {\bibinfo {author} {\bibfnamefont {D.}~\bibnamefont
  {Giovannini}}, \bibinfo {author} {\bibfnamefont {F.~M.}\ \bibnamefont
  {Miatto}}, \bibinfo {author} {\bibfnamefont {J.}~\bibnamefont {Romero}},
  \bibinfo {author} {\bibfnamefont {S.~M.}\ \bibnamefont {Barnett}}, \bibinfo
  {author} {\bibfnamefont {J.~P.}\ \bibnamefont {Woerdman}}, \ and\ \bibinfo
  {author} {\bibfnamefont {M.~J.}\ \bibnamefont {Padgett}},\ }\href
  {http://stacks.iop.org/1367-2630/14/i=7/a=073046} {\bibfield  {journal}
  {\bibinfo  {journal} {New Journal of Physics}\ }\textbf {\bibinfo {volume}
  {14}},\ \bibinfo {pages} {073046} (\bibinfo {year} {2012})}\BibitemShut
  {NoStop}%
\bibitem [{\citenamefont {Vasnetsov}\ \emph {et~al.}(2003)\citenamefont
  {Vasnetsov}, \citenamefont {Torres}, \citenamefont {Petrov},\ and\
  \citenamefont {Torner}}]{Vasnetsov:03}%
  \BibitemOpen
  \bibfield  {author} {\bibinfo {author} {\bibfnamefont {M.~V.}\ \bibnamefont
  {Vasnetsov}}, \bibinfo {author} {\bibfnamefont {J.~P.}\ \bibnamefont
  {Torres}}, \bibinfo {author} {\bibfnamefont {D.~V.}\ \bibnamefont {Petrov}},
  \ and\ \bibinfo {author} {\bibfnamefont {L.}~\bibnamefont {Torner}},\ }\href
  {\doibase 10.1364/OL.28.002285} {\bibfield  {journal} {\bibinfo  {journal}
  {Opt. Lett.}\ }\textbf {\bibinfo {volume} {28}},\ \bibinfo {pages} {2285}
  (\bibinfo {year} {2003})}\BibitemShut {NoStop}%
\bibitem [{\citenamefont {Berkhout}\ and\ \citenamefont
  {Beijersbergen}(2008)}]{Berkhout:08}%
  \BibitemOpen
  \bibfield  {author} {\bibinfo {author} {\bibfnamefont {G.~C.~G.}\
  \bibnamefont {Berkhout}}\ and\ \bibinfo {author} {\bibfnamefont {M.~W.}\
  \bibnamefont {Beijersbergen}},\ }\href {\doibase
  10.1103/PhysRevLett.101.100801} {\bibfield  {journal} {\bibinfo  {journal}
  {Phys. Rev. Lett.}\ }\textbf {\bibinfo {volume} {101}},\ \bibinfo {pages}
  {100801} (\bibinfo {year} {2008})}\BibitemShut {NoStop}%
\bibitem [{\citenamefont {Lavery}\ \emph {et~al.}(2011)\citenamefont {Lavery},
  \citenamefont {Berkhout}, \citenamefont {Courtial},\ and\ \citenamefont
  {Padgett}}]{Lavery:11}%
  \BibitemOpen
  \bibfield  {author} {\bibinfo {author} {\bibfnamefont {M.~P.~J.}\
  \bibnamefont {Lavery}}, \bibinfo {author} {\bibfnamefont {G.~C.~G.}\
  \bibnamefont {Berkhout}}, \bibinfo {author} {\bibfnamefont {J.}~\bibnamefont
  {Courtial}}, \ and\ \bibinfo {author} {\bibfnamefont {M.~J.}\ \bibnamefont
  {Padgett}},\ }\href {http://stacks.iop.org/2040-8986/13/i=6/a=064006}
  {\bibfield  {journal} {\bibinfo  {journal} {Journal of Optics}\ }\textbf
  {\bibinfo {volume} {13}},\ \bibinfo {pages} {064006} (\bibinfo {year}
  {2011})}\BibitemShut {NoStop}%
\bibitem [{\citenamefont {Mair}\ \emph {et~al.}(2001)\citenamefont {Mair},
  \citenamefont {Vaziri}, \citenamefont {Weihs},\ and\ \citenamefont
  {Zeilinger}}]{Mair:01}%
  \BibitemOpen
  \bibfield  {author} {\bibinfo {author} {\bibfnamefont {A.}~\bibnamefont
  {Mair}}, \bibinfo {author} {\bibfnamefont {A.}~\bibnamefont {Vaziri}},
  \bibinfo {author} {\bibfnamefont {G.}~\bibnamefont {Weihs}}, \ and\ \bibinfo
  {author} {\bibfnamefont {A.}~\bibnamefont {Zeilinger}},\ }\href {\doibase
  10.1038/35085529} {\bibfield  {journal} {\bibinfo  {journal} {Nature}\
  }\textbf {\bibinfo {volume} {412}},\ \bibinfo {pages} {313} (\bibinfo {year}
  {2001})}\BibitemShut {NoStop}%
\bibitem [{\citenamefont {Qassim}\ \emph {et~al.}(2014)\citenamefont {Qassim},
  \citenamefont {Miatto}, \citenamefont {Torres}, \citenamefont {Padgett},
  \citenamefont {Karimi},\ and\ \citenamefont {Boyd}}]{Qassim:14}%
  \BibitemOpen
  \bibfield  {author} {\bibinfo {author} {\bibfnamefont {H.}~\bibnamefont
  {Qassim}}, \bibinfo {author} {\bibfnamefont {F.~M.}\ \bibnamefont {Miatto}},
  \bibinfo {author} {\bibfnamefont {J.~P.}\ \bibnamefont {Torres}}, \bibinfo
  {author} {\bibfnamefont {M.~J.}\ \bibnamefont {Padgett}}, \bibinfo {author}
  {\bibfnamefont {E.}~\bibnamefont {Karimi}}, \ and\ \bibinfo {author}
  {\bibfnamefont {R.~W.}\ \bibnamefont {Boyd}},\ }\href {\doibase
  10.1364/JOSAB.31.000A20} {\bibfield  {journal} {\bibinfo  {journal} {J. Opt.
  Soc. Am. B}\ }\textbf {\bibinfo {volume} {31}},\ \bibinfo {pages} {A20}
  (\bibinfo {year} {2014})}\BibitemShut {NoStop}%
\bibitem [{\citenamefont {Riza}(2013)}]{Riza:13}%
  \BibitemOpen
  \bibfield  {author} {\bibinfo {author} {\bibfnamefont {N.}~\bibnamefont
  {Riza}},\ }\href@noop {} {\emph {\bibinfo {title} {Photonic Signals and
  Systems: An Introduction}}}\ (\bibinfo  {publisher} {McGraw Hill},\ \bibinfo
  {year} {2013})\BibitemShut {NoStop}%
\bibitem [{\citenamefont {Sheikh}\ and\ \citenamefont
  {Rehman}(2015)}]{Sheikh:15}%
  \BibitemOpen
  \bibfield  {author} {\bibinfo {author} {\bibfnamefont {M.}~\bibnamefont
  {Sheikh}}\ and\ \bibinfo {author} {\bibfnamefont {S.~A.}\ \bibnamefont
  {Rehman}},\ }in\ \href@noop {} {\emph {\bibinfo {booktitle} {Third
  International Conference on Optical Angular Momentum}}}\ (\bibinfo {year}
  {2015})\BibitemShut {NoStop}%
\bibitem [{\citenamefont {Peatross}\ and\ \citenamefont
  {Ware}(2015)}]{Peatross:15}%
  \BibitemOpen
  \bibfield  {author} {\bibinfo {author} {\bibfnamefont {J.}~\bibnamefont
  {Peatross}}\ and\ \bibinfo {author} {\bibfnamefont {M.}~\bibnamefont
  {Ware}},\ }\href@noop {} {\emph {\bibinfo {title} {Physics of Light and
  Optics}}}\ (\bibinfo  {publisher} {optics.byu.edu},\ \bibinfo {year}
  {2015})\BibitemShut {NoStop}%
\bibitem [{\citenamefont {Goodman}(2005)}]{Goodman:05}%
  \BibitemOpen
  \bibfield  {author} {\bibinfo {author} {\bibfnamefont {J.}~\bibnamefont
  {Goodman}},\ }\href@noop {} {\emph {\bibinfo {title} {Introduction to Fourier
  Optics}}},\ \bibinfo {edition} {3rd}\ ed.\ (\bibinfo  {publisher} {Roberts
  and Company},\ \bibinfo {year} {2005})\BibitemShut {NoStop}%
\bibitem [{\citenamefont {Molina-Terriza}\ \emph {et~al.}(2007)\citenamefont
  {Molina-Terriza}, \citenamefont {Rebane}, \citenamefont {Torres},
  \citenamefont {Torner},\ and\ \citenamefont {Carrasco}}]{Molina-Terriza:07}%
  \BibitemOpen
  \bibfield  {author} {\bibinfo {author} {\bibfnamefont {G.}~\bibnamefont
  {Molina-Terriza}}, \bibinfo {author} {\bibfnamefont {L.}~\bibnamefont
  {Rebane}}, \bibinfo {author} {\bibfnamefont {J.}~\bibnamefont {Torres}},
  \bibinfo {author} {\bibfnamefont {L.}~\bibnamefont {Torner}}, \ and\ \bibinfo
  {author} {\bibfnamefont {S.}~\bibnamefont {Carrasco}},\ }\href
  {http://www.jeos.org/index.php/jeos_rp/article/view/07014} {\bibfield
  {journal} {\bibinfo  {journal} {Journal of the European Optical Society -
  Rapid publications}\ }\textbf {\bibinfo {volume} {2}} (\bibinfo {year}
  {2007})}\BibitemShut {NoStop}%
\bibitem [{\citenamefont {Kogelnik}\ and\ \citenamefont
  {Li}(1966)}]{Kogelnik:66}%
  \BibitemOpen
  \bibfield  {author} {\bibinfo {author} {\bibfnamefont {H.}~\bibnamefont
  {Kogelnik}}\ and\ \bibinfo {author} {\bibfnamefont {T.}~\bibnamefont {Li}},\
  }\href {\doibase 10.1364/AO.5.001550} {\bibfield  {journal} {\bibinfo
  {journal} {Appl. Opt.}\ }\textbf {\bibinfo {volume} {5}},\ \bibinfo {pages}
  {1550} (\bibinfo {year} {1966})}\BibitemShut {NoStop}%
\bibitem [{\citenamefont {Qasim}\ and\ \citenamefont {Reza}(2015)}]{Qasim:15}%
  \BibitemOpen
  \bibfield  {author} {\bibinfo {author} {\bibfnamefont {M.}~\bibnamefont
  {Qasim}}\ and\ \bibinfo {author} {\bibfnamefont {S.~A.}\ \bibnamefont
  {Reza}},\ }\href {\doibase 10.1364/AO.54.009242} {\bibfield  {journal}
  {\bibinfo  {journal} {Appl. Opt.}\ }\textbf {\bibinfo {volume} {54}},\
  \bibinfo {pages} {9242} (\bibinfo {year} {2015})}\BibitemShut {NoStop}%
\end{thebibliography}%

\end{document}